\documentclass{sig-alternate-05-2015}

\usepackage{graphicx}
\usepackage{color} 
\usepackage{subfigure}

\begin{document}
%

\title{Can NDN Perform Better than OLSR in Wireless Ad Hoc Networks?}
%
%
%
%
%

\numberofauthors{1} 
%
\author{
%
%
\alignauthor
Thiago Teixeira, Cong Wang, Michael Zink\\
       \affaddr{University of Massachusetts Amherst}\\
       \affaddr{151 Holdsworth Way}\\
       \affaddr{Amherst, Massachusetts}\\
       \email{\{tteixeira, cwang1, mzink\}@umass.edu}
\alignauthor
}

\maketitle
\begin{abstract}
The emerging paradigm of Information-Centric Networking is an exciting field of research, opening opportunities in many areas, such as forwarding strategies, caching placement policies, applications (e.g. video streaming and instant messaging), to name a few. In this paper, we address the mobility aspect of ICN, as well as how it performs in tactical wireless ad hoc environments. In this paper, we present results from a simulation study that investigates the performance of Named Data Networking, an instantiation of ICN, in such environments.
We perform a series of simulations based on ndnSIM studying different mobility scenarios.
Our simulations show that even in the short-term absence of the producer, consumers can still achieve better file retrieval when caches are used. As an effort to increase the cache diversity and have a better utilization of the Content Store we study probabilistic LRU. 
Furthermore, we compare the performance of our NDN network with a TCP based approach, using OLSR routing protocol, discussing advantages and disadvantages of each approach. 

\end{abstract}




\section{Introduction}
\label{sec:intro}
The Internet is constantly evolving, from circuit-switching to packet-switching, to content distribution, which cope with the increasingly amount of content transferred over the Internet. In recent years, several proposals for a new Internet architecture have arisen. Among the most successful concepts is Information-Centric Networking (ICN), which promotes a profound change by reusing some of TCP/IP's successful ideas while fixing its shortcomings. One of the most significant differences in ICN is the addressing of content by name instead of an address. This is, e.g., the case in Named Data Networking (NDN) \cite{Jacobson:2009:NNC:1658939.1658941} -- a specific architecture design of ICN.

Initially, content is located on the producer \footnote{We use the terms producer, custodian, and source interchangeably from now on to refer to the content producer} end, where it is permanently stored. The consumer initiates Interest requests, which are forwarded via NDN routers towards the producer. In contrast to IP routers their NDN equivalents contain a Pending Interest Table (PIT) and a Content Store (CS)
in addition to the Forwarding Information Base. 
When an NDN router forwards an Interest, it creates an entry in the PIT and forwards the Interest based on the FIB. When the router receives the Data packet, it removes the PIT entry, forwards the Data to the appropriate faces, and store a copy of the Data in the CS. Therefore, whenever a different node requests the same Data, any router that has the content stored can serve the request.

This type of pervasive in-network caching is an interesting feature for  mobile ad hoc networks (MANET), as nodes frequently move in and out of communication range. Any node in possession of the requested content can serve the request without the need of retrieving it from the source. We are interested in a special case of MANETs where the mobile nodes have scarce resources, e.g., battery operated radios with limited transmission power, resulting in a lower communication range. Networks with this characteristics are called disconnected, intermittent, low-bandwidth (DIL).

In this paper, we compare the performance of NDN with Optimized Link-State Routing (OLSR) \cite{Clausen:2003:OLS:RFC3626}, a wireless ad hoc routing protocol designed to reduce the flooding of messages in the wireless media by selecting specific nodes, called multipoint relays (MPRs), to re-broadcast packets.

We simulate both protocols in a mesh network using NS-3 \cite{ns3} and ndnSIM \cite{399} for OLSR and NDN, respectively. 
The main goal of the simulations is to evaluate the strengths and weaknesses of NDN in comparison to OLSR in a MANET scenario. We specifically simulate cases of network partitioning and random mobility, since both represent important characteristics of MANETs. In the case of 
network partitioning, we investigate if the intrinsic caching characteristics of NDN will impact overall performance in cases where the consumer can not communicate with the producer. In the random mobility case, we investigate if NDN's in-network caching will improve performance compared to an OLSR-based approach.


We further study the caching strategies in the NDN routers, aiming to enhance the data delivery and cache utilization. We make use of probabilistic least recent used (PLRU) caching algorithm.

The main contributions of this paper are the following:

\begin{itemize}
\item We show that NDN performance in wireless ad hoc networks still needs improvements if we compare to OLSR.

\item We evaluate probabilistic caching as a strategy in NDN routers.

\item Broadens the discussion of NDN forwarding strategies for wireless ad hoc networks.
\end{itemize} 

The remainder of this paper is organized as follows: Section \ref{Motivation} introduces our research goals, Section \ref{sec:related} discusses related work, Sections \ref{sec:icn-support} present the simulation scenarios. We present an evaluation of our simulation cases in Section \ref{sec:evaluation}. Section \ref{Conclusion} summarizes and concludes the paper.

\section{Motivation} \label{Motivation}
In MANETs, nodes are moving and may go in and out of communication range, making it burdensome to establish a end-to-end path. TCP/IP based protocols were not initially designed to handle this problem. The existing approaches rely on a rendezvous mechanism to track node mobility \cite{zhangsurvey} in order to move data towards the receiver. Another problem in wireless ad hoc networks is the shared media. Existing protocols reduce the probability of collisions by scheduling transmission, reducing the number of broadcast nodes, or a combination of both, to name a few.

The pervasive in-network caching in NDN, may makes it easier to move content towards the receiver while reducing the probability of collisions in the network.
This reduction is caused by the "merging" of requests in NDN routers. E.g., when a NDN router receives a new Interest, it checks its PIT and if the Interest is already 
pending, the router does not forward the Interest, reducing traffic in the network. Also, NDN routers check if the requested Data is in its Content Store. Cached content 
will also reduce collisions since it as not to be sent all the way from the source of the content.

Our hypothesis is that because NDN routers suppress broadcasts in the aforementioned cases, it will outperform OLSR in our simulated network.

\section{Related Work}
\label{sec:related}
Previous works have compared the performance of ICNs with TCP/IP-based networks and described caching placement policies as well as forwarding strategies to increase performance in the network. In this section, we briefly describe the work performed in these areas.

\subsection{Information-Centric Networking}
In \cite{Lertsinsrubtavee:2014:CNC:2684793.2684800}, authors have performed an evaluation of Named-Data Networking and Content Distribution Networks (CDNs) in a rural community wireless mesh network (CWMN). Their results show that the CDN approach achieves better performance in terms of content delivery time; however, due to the in-network caching, NDN generated less cache misses, leading to a lower bandwidth usage.

A similar evaluation was done in \cite{Teixeira:2016:EIN:2881025.2889483}, where the authors showed that NDN outperforms TCP in a wireless ad hoc network when the hop count is high. TCP congestion control does not perform well in high hop count scenarios where the probability of collisions is high, 
as every packet loss event will negatively impact the congestion window.

In \cite{Amadeo2015148} and \cite{Wang:2012:RTI:2248361.2248365} authors have proposed different strategies for forwarding in wireless ad hoc networks.

Amadeo et al. \cite{Amadeo2015148} investigates the trade-off between blind forwarding (BF) and provider-aware forwarding (PAF) in wireless ad hoc networks. The first approach does not take into account the neighboring nodes but a timer-based forwarding strategy to re-broadcast packets. Both Interest and Data packets are randomly rebroadcasted, taking into account the defer-window (DW), an integer value indicating the length of rebroadcast interval. We make use this forwarding strategy in our paper. PAF on the other hand, keeps track of the neighboring nodes by using a distance table and the hop distance to the content provider. When sending the next Interest, the consumer includes this information in the request so intermediate nodes can refrain from rebroadcasting. PAF is an extension of BF, with the expense of a higher control overhead.

Wang et al. \cite{Wang:2012:RTI:2248361.2248365} address the dissemination of information on a highway by implementing Vehicle-to-Vehicle communication using NDN. This approach proactively pushes data to possible consumers through data mules, making use of timers to control collisions in the wireless media. As the producer and consumers are not known a priori, NDN showed better performance compared to IP-based approaches.

\subsection{Caching}
Cache placement policies are investigated in \cite{Badov:2014:CCS:2660129.2660145} where the authors found that caching in the downstream end of a low capacity or congested link makes the most effective use of cache space. Furthermore, caching policies targeted to achieve the best network-centric performance may provide poor user-centric performance. Finally, the authors developed a congestion-aware cache management policy, based on an utility function that estimates local congestion, content popularity, and end-to-end throughput.

In an effort to improve delivery time and reduce redundancy of cached contents, Psaras et al. have explored probabilistic caching \cite{Psaras:2012:PIC:2342488.2342501}. The authors have developed ProbCache, which approximates the cache probability of a path, increasing the caching probability as Data travels away from the custodian. Results showed up to a 20\% reduction in server hits and up to 10\% reduction in hop count, when compared with universal caching schemes developed for web caching.

In our approach, we change the caching probability of the nodes closer to the consumers, adding diversity in the consumer end of the network as we are interested in benefiting the secondary consumers that request the same content chunks at a later time.

\subsection{Mobility}
A study on the existing mobility solutions is done in \cite{zhangsurvey}. Zhang et al. grouped the IP mobility patterns in routing-based, mapping-based, or tracing-based, and the NDN mobility in consumer mobility and producer mobility.

Also, Meisel et al. \cite{Meisel:2010:AHN:1859983.1859986} evaluated node mobility in ad hoc networks using Listen First, Broadcast Later (LFBL) protocol for NDN nodes and AODV protocol for TCP/IP nodes. NDN showed better performance when nodes are mobile, while performance degrades in the AODV cases.

Azgin et al.~\cite{6883822} studied the NDN support for mobility in wireless access networks, focusing on consumer and producer mobility within the same Autonomous System (AS) (i.e. intra-AS), as well as multiple ASes (i.e., inter-AS). The unmodified NDN architecture, simulated using ndnSIM, presented significant user experience degradation when mobility is introduced; therefore, raising the need for developing location-centric routing mechanisms.

A performance comparison between IP and NDN MANETs in natural disaster scenarios has been performed by Alubady et al. \cite{alubadyinternet}. The authors investigated the performance of both protocols when the node count and packet size changes. NDN outperformed the so-called IP-MANET using destination sequenced distance vector (DSDV) routing algorithm.

While \cite{alubadyinternet} compares a NDN MANET and an IP-based MANET varying node density and packet size, we compare both protocols (though using OLSR for our IP-based MANET) when mobility and network fragmentation is used. Moreover, the network topology and other parameters are not quite explicit in their paper.

\section{ICN support for ad hoc} 
\label{sec:icn-support}
Information dissemination in wireless ad hoc networks is challenging in the sense that nodes move in and out of communication range at any given point in time and communicate without a central controller. As we discussed previously, most solutions use a random timer to schedule transmission of packets, in an effort to reduce collisions.

Amadeo et al. \cite{Amadeo2015148} developed controlled flooding forwarding strategy for NDN, where nodes listen to the channel for a random time, refraining from broadcasting if the Interest or Data packet in its forwarding queue is overheard in the channel. The current NDN simulator (ndnSIM \cite{399}) does not perform well in multi-hop wireless ad hoc networks by default \cite{6883822}, \cite{Amadeo2015148}. The reason for that is because the original design does not allow an Interest request or data packet to be forwarded in the same face that the Interests/Data packet was received. In the wired scenario, it prevents loops from happening, but in the wireless case it is necessary. For this reason, we use controlled flooding in our simulations.

Because of in-network pervasive caching in ICNs is an interesting feature for wireless ad hoc networks, our hypothesis is that NDN will outperform the current TCP/IP based protocols, such as OLSR when the network experience fragmentation. In the following subsections we describe our experiments to test our hypothesis.

\subsection{NDN vs. OLSR} \label{NDN vs. OLSR}
To asses the performance of our mesh network using NDN and OLSR we developed simulations with controlled mobility (where nodes move according to a certain pattern) as well as random mobility (where nodes move in a Brownian motion fashion).

Table \ref{tab:configurations} shows the simulation parameters we use for the evaluation. We define the file size to be downloaded by the consumers to be approximately 1 MByte, divided in a thousand packets. The wireless media consists of a 1 Mbps channel. Each node is equipped with an IEEE 802.11b interface with DSSS modulation. The remaining wireless parameters configure a communication range of just over a 100 meters (distance between nodes in our grid). 
The generation of Interests by the consumers is based on round-trip time estimation to avoid flooding of the network and adapt to current conditions in the network.

Finally, we use the defer window value (explained previously) of 127. We found, in previous simulations, that this value provides a good balance between overhearing transmissions in the wireless channel without holding the packets for too long.

The following subsections describe network topology and mobility patterns.

\begin{table}
\centering
\caption{Simulation parameters}
\resizebox{.5\textwidth}{!}{
\begin{tabular}{c c c c} \hline
Category & Parameter & TCP/UDP & ICN\\ \hline
Application layer & File size & \multicolumn{2}{c}{1000 packets}\\ 
 & Data payload & \multicolumn{2}{c}{1040 bytes}\\ 
In-network caching & Content store size& - & 100 \\ 
Wireless media & Propagation model & \multicolumn{2}{c}{Friis} \\ 
 & Protocol &  \multicolumn{2}{c}{IEEE 802.11b DSSS} \\
 & Channel capacity & \multicolumn{2}{c}{1 Mbps} \\
 & Rx sensitivity & \multicolumn{2}{c}{-80 dBm} \\
 & Tx power & \multicolumn{2}{c}{5 dBm} \\
Node layout & Grid topology & \multicolumn{2}{c} {21 nodes}\\ 
& Random & \multicolumn{2}{c}{Random walk}\\
& Velocity & \multicolumn{2}{c}{6 m/s} \\ 
Consumer Application &  & On/Off & Batches \\ 
Transport Layer & & TCP, UDP & NDN\\ 
NDN defer window size & & N/A & 127 \\
\hline
\end{tabular}
} 
\label{tab:configurations}%
\end{table}

\subsubsection{Controlled mobility} \label{Controlled mobility}
\begin{figure}[!t]
\centering
\includegraphics[height=1.5in, width=3.3in]{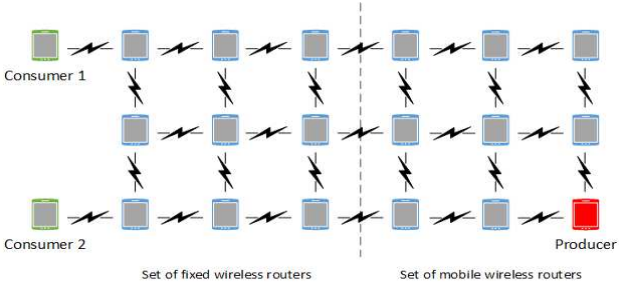}
\caption{Network topology for controlled mobility scenario}
\label{fig:controlled mobility}
\end{figure}

To better understand how different protocols respond to changes in the network traffic and how producer mobility affects file retrieval, we decided to 
choose the topology shown in Figure \ref{fig:controlled mobility} for our simulations.

The topology is comprised of two consumers, two sets of wireless nodes 	spaced 100 meters apart from each other, and one producer. One set of wireless routers is labeled as fixed in Figure \ref{fig:controlled mobility}. These routers do not move during the simulation. The other set of wireless routers is labeled as mobile. These nodes, including the producer, move away in a convoy fashion during the simulation while consumers are retrieving the file. After a disconnected period, the mobile nodes move back to their original position. The goal of this particular simulation is to investigate how OLSR and NDN perform in the case of network partition.

We start consumer 1 at t = 0 s, and vary the time at which consumer 2 starts (e.g. 5, 10, 50, 100, and 200 s). The reason for that is that we want to understand by how much caches play a role in the file retrieval, as well as network traffic. For the NDN application, both consumers request content chunks in the same namespace (i.e. /test/content).

In the NDN case, requests for the same content are merged by the routers, using less resources in terms of channel capacity and energy. Also, when the consumer sends an Interest request, it starts a timer. Eventually this request will reach the content source and the respective Content is sent back to the consumer. In case of a collision, which prevents the Content from traveling further towards the consumer, the Content is still cached at the previous hop. Thus, when the consumers initiates a new Interest request (e.g., after a time out), it will most likely find that Content cached in the network. We can observe this pattern in our simulations and will present results from our simulations in Section \ref{sec:evaluation}.

\subsubsection{Random mobility} \label{Random mobility}
\begin{figure}[!t]
\centering
\includegraphics[height=2.0in, width=2.0in]{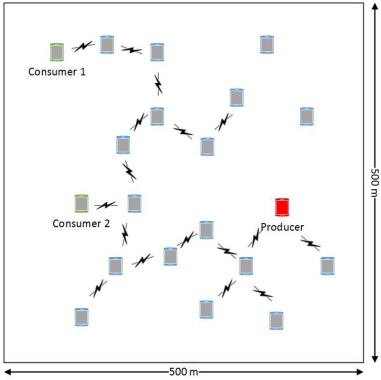}
\caption{Network topology for the random mobility scenario at start time}
\label{fig:random mobility}
\end{figure}

In order to understand how different protocols perform in random mobility scenarios (e.g. soldiers in a battlefield carrying mobile transceivers), we simulate
a scenario in which node are placed at random positions within a 500 x 500 meters area.
At the start of the simulation, all nodes start to move in a random fashion according to the Random Walk 2D Mobility Model.
When a node reaches the boundary of the simulated area it rebounds with a reflexive angle and speed.
All other simulation parameters remain the same (i.e. same number of nodes, same wireless channel, same Tx/Rx parameters, see Section~\ref{NDN vs. OLSR}). The topology for this scenario is depicted in Figure \ref{fig:random mobility}. Note that at start time, some nodes may not be in range of communication of others (no communication link).

We start consumer 1 at t = 0 s, and vary the time at which consumer 2 starts (e.g., 0, 10, and 100 s). The reasoning behind that is two fold. First, we want to understand how consumers can retrieve files in intermittent and disconnected environments. Second, we want to understand how caches benefit the overall network performance. We study an alternative caching strategy in the next subsection.

\subsubsection{Caching} \label{Caching}
In addition to the popular least recently used (LRU) cache replacement policy, in which caches are refreshed with the newest content and the oldest content (least used in this case) are evicted, we also investigate how probabilistic LRU (pLRU) impacts the overall system performance of an NDN-based MANET.
In the specific case of our simulations caches are completely refreshed after approximately 10 seconds if LRU is applied and the content of caches will be fairly similar.

One way to increase cache diversity is the usage of pLRU. In our simulation we set $p=0.5$, which basically means in case a node receives a Content chunk, it flips a coin and caches the content or not.
In the controlled mobility scenario, we changed the caching probability in the fixed nodes only. The reason for that is that we want to guarantee that in case of collisions, the content is cached as further away as possible from the custodian. In the random mobility scenario, we changed the caching probability in all mobile nodes, since all of them are randomly placed.

In the following section, we present and analyze our results.

\section{Evaluation} 
\label{sec:evaluation}
In this section, we present a performance analysis of the aforementioned simulation scenarios. 
We run each simulation scenario with five different seeds and present the average, maximum, and minimum values.

When comparing the NDN scenarios, we use the metrics file retrieval delay, hop count, and number of retransmissions. Each metric is 
explained in more detail in the following.

\begin{itemize}
\item Content retrieval delay: the time elapsed from the first Interest request issued for a specific content to the time that specific content is received, accounting for all eventual retransmissions. It measures how fast the network can transfer data.

\item Hop count: is the number of hops that a Data packet has to travel from the content source (i.e. custodian or network cache) to the content sink (i.e. consumer). The hop count metric only accounts for the last successful transmission.

\item Retransmissions: because PITs have limited resources, a timer is started each time an Interest request is sent out by one of the consumers. This timer should be greater than the RTT to the source (calculated by Jacobson's estimation); however, if an Interest request or Data packet gets lost in the network, consumers have to issue another Interest request, increasing the retransmission count. Please note that the first transmission is accounted in Figure \ref{fig:NDN Metrics} (c).

\end{itemize} 

Because the studied protocols have different reliability mechanisms and different packet overheads, we compare their performance using goodput and packet loss rate, described as follows.

\begin{itemize}
\item Goodput: is the total delivered payload (in Bytes) from the time at which the first packet is transmitted to the time at which the last packet is received, depicted in Equation \ref{eq:goodput}. We believe goodput provides a more fair comparison to evaluate the protocols as it accounts for the effective throughput.

\begin{equation} \label{eq:goodput}
Goodput = \frac{Rx Bytes * 8}{TimeLastRxPacket - TimeFirstTxPacket}
\end{equation} 

\item Packet loss: packet loss refers to the rate of total packets lost and the total packets sent during transmission. Packet loss is illustrated in Equation \ref{eq:lossrate}, where $p_{loss}$ is the total number of lost packets, and $p_{sent}$ is the total number of packets sent.

\begin{equation} \label{eq:lossrate}
Loss Rate = \frac{p_{loss}}{p_{sent}}
\end{equation} 

\end{itemize}

\subsection{Controlled mobility}
\begin{figure*}[t]
 \centering
   \subfigure[]{\includegraphics[width=0.32\textwidth]{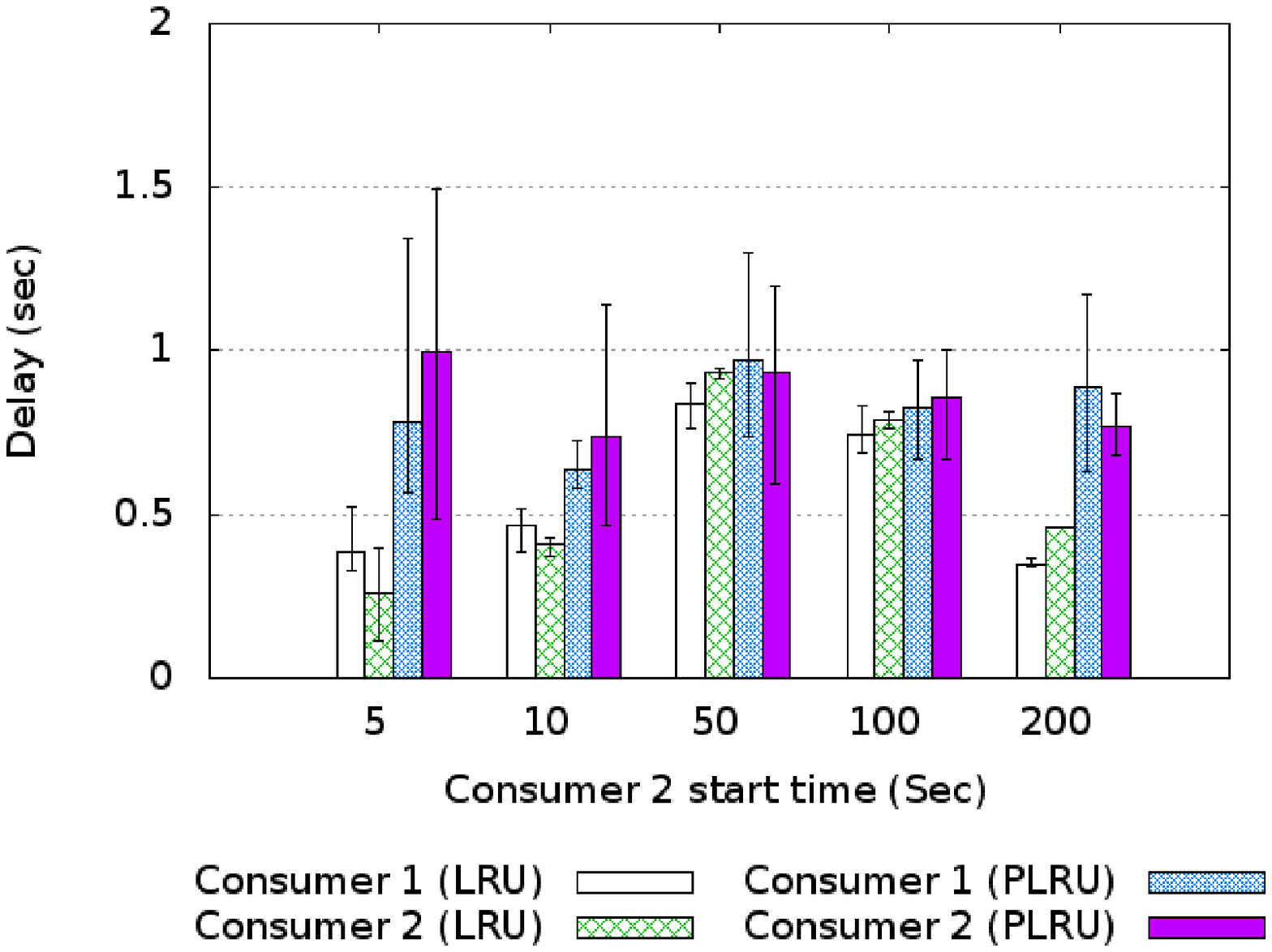}\label{fig:NDN_delay}}
   \subfigure[]{\includegraphics[width =0.32\textwidth]{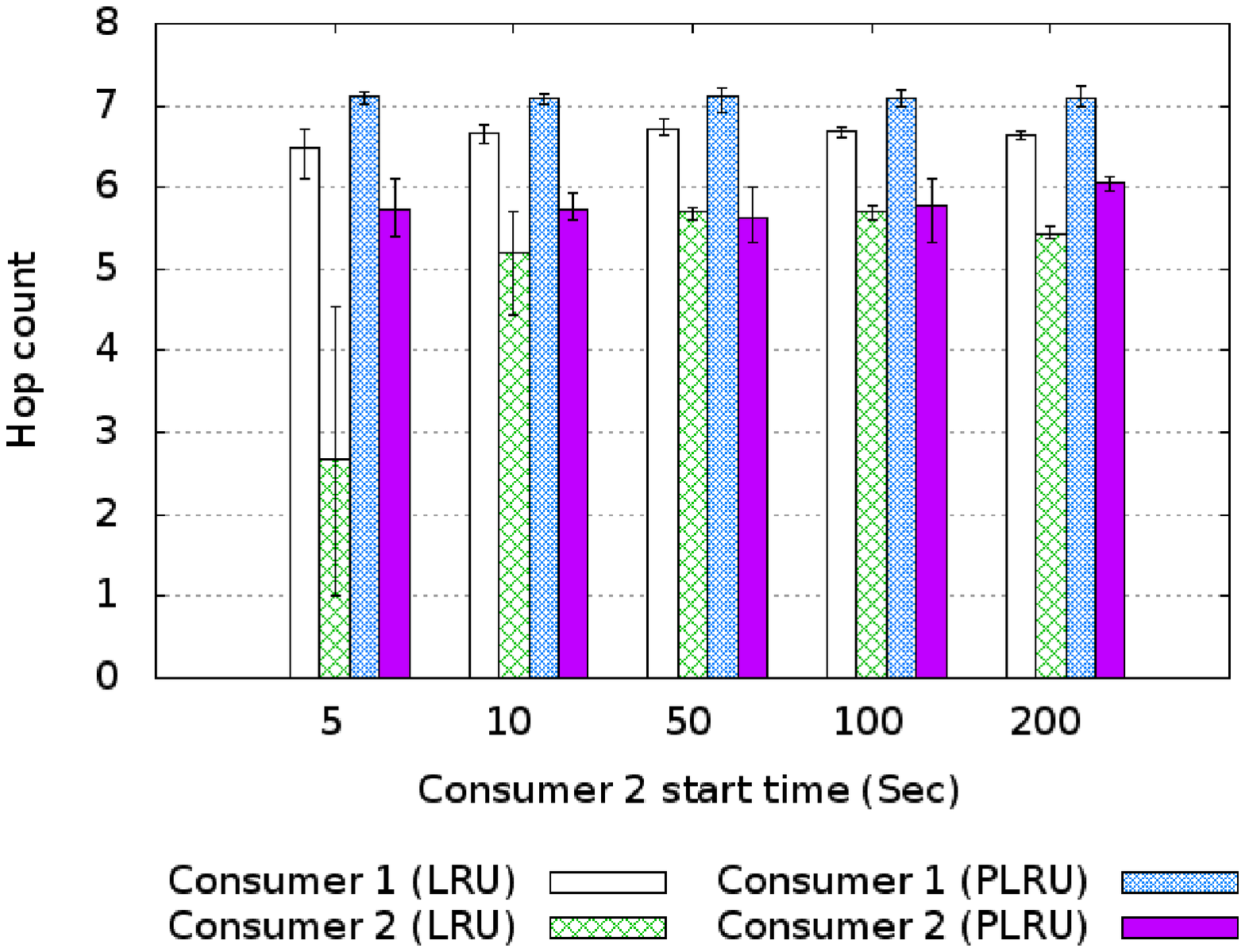}\label{fig:NDN_hopcout}}
   \subfigure[]{\includegraphics[width =0.32\textwidth]{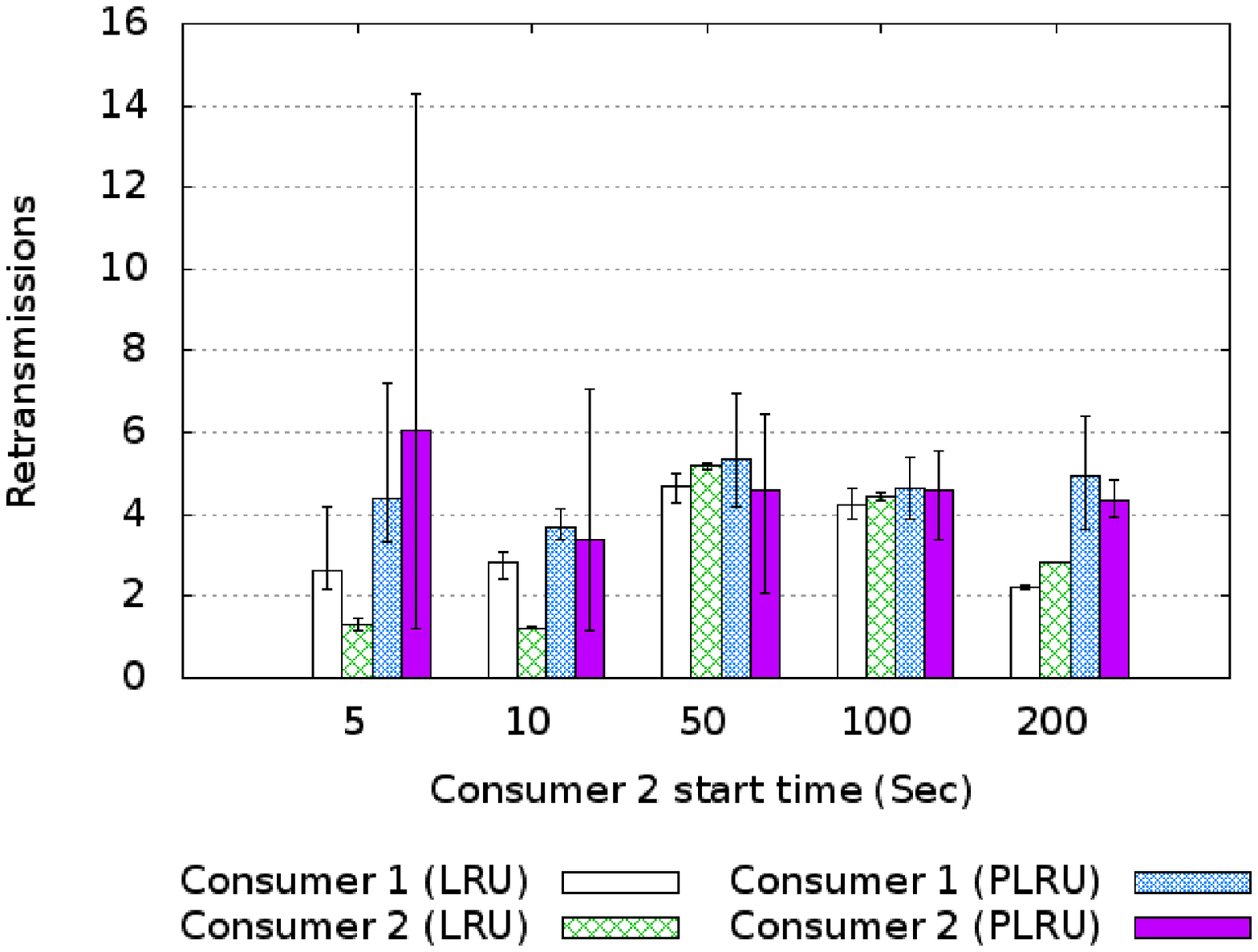}\label{fig:NDN_retrans}}
\caption{NDN - Data retrieval delay, hop count, and retransmission count for consumers 1 and 2 when we vary the time which consumer 2 starts.}\label{fig:NDN Metrics}
 \end{figure*}

We begin our analysis by initially discussing the results for LRU shown in Figure \ref{fig:NDN Metrics} (a). The results for PLRU will be discussed later. For the case where Consumer 2 starts at t=5 seconds, we notice that Consumer 2 finds the requested content cached in the network, as Consumer 1 has already requested them and they are stored in the content stores of the intermediated nodes. This translates to a lower Content retrieval delay and is corroborated by the lower hop count and number of retransmissions as shown in Figure \ref{fig:NDN Metrics} (b) and (c), respectively.

A similar analysis can be drawn for the case where Consumer 2 starts at t=10 seconds. In this case, some caches in the network have already been refreshed with newer content, pushing the average retrieval delay for Consumer 2 up, as well as the retrieval delay for Consumer 1, as seen in Figure \ref{fig:NDN Metrics} (a). This is because the traffic in the network has been increased and Consumer 2 has to retrieve content from nodes that are farther away (higher hop count seen in Figure \ref{fig:NDN Metrics} (b)).

Note that  although the custodian moves away for a short amount of time during the simulation in the previous two cases, the overall network performance does not experience much degradation.

When Consumer 2 starts at t=50 seconds, all caches are already refreshed, since Consumer 1 has been downloading the file for 50 seconds. In this case, both consumers are downloading different content chunks at the same time, increasing the traffic in the network. This is why the average retrieval delay and retransmission count in this case are the highest among all. 

As previously mentioned in Section \ref{Controlled mobility}, the lower hop count and higher number of retransmissions in Figures \ref{fig:NDN Metrics} (b) and (c) may seems counter intuitive at first. When a consumer sends out an Interest request and it reached the custodian, the Data packet starts to travel back, following the breadcrumbs and being cached along the path. In case the timer at the consumer expires, the consumer expresses another Interest request for that same content; however, it will most likely find that content cached in an intermediate node. Thus, the lower hop count and higher retransmission rate.

For the case where Consumer 2 starts at t=100 s, we find the network less congested than the previous case, as seen in Figure \ref{fig:NDN Metrics} (a). This is because there is less overlap between transmission from Consumer 1 and 2. The remaining parameters are similar to the previous case. 

Finally, the case where Consumer 2 starts at t=200 seconds is when the network is less congested overall, as can be seen in Figure \ref{fig:NDN Metrics} (a). Since caches are refreshed when Consumer 2 starts downloading the file we do not see significant changes in the hop count, compared to the previous cases. Though the retransmission count in this case is affected by the network partition, specially for Consumer 2, as seen in Figure \ref{fig:NDN Metrics} (c).

For the probabilistic caching cases, the expectation is that the download time would decrease when caches are refreshed (i.e. 50, 100, and 200 cases), since Consumer 2 would find some contents cached in the network. Although PLRU achieved very similar results to LRU in the 50 and 100 second cases (in some cases better), on average it does not perform as well as pure LRU. In addition, the PLRU results show a higher variance for all three metrics.

In general, changing the caching probability in the wireless routers did not show significant improvements in performance, on average. Therefore, we need further study probabilistic caching in wireless ad hoc networks.

\begin{figure*}[!t]
\centering
  \begin{tabular}{@{}cc@{}}
	(a) & (b)  \\   
    \includegraphics[width=.3\textwidth]{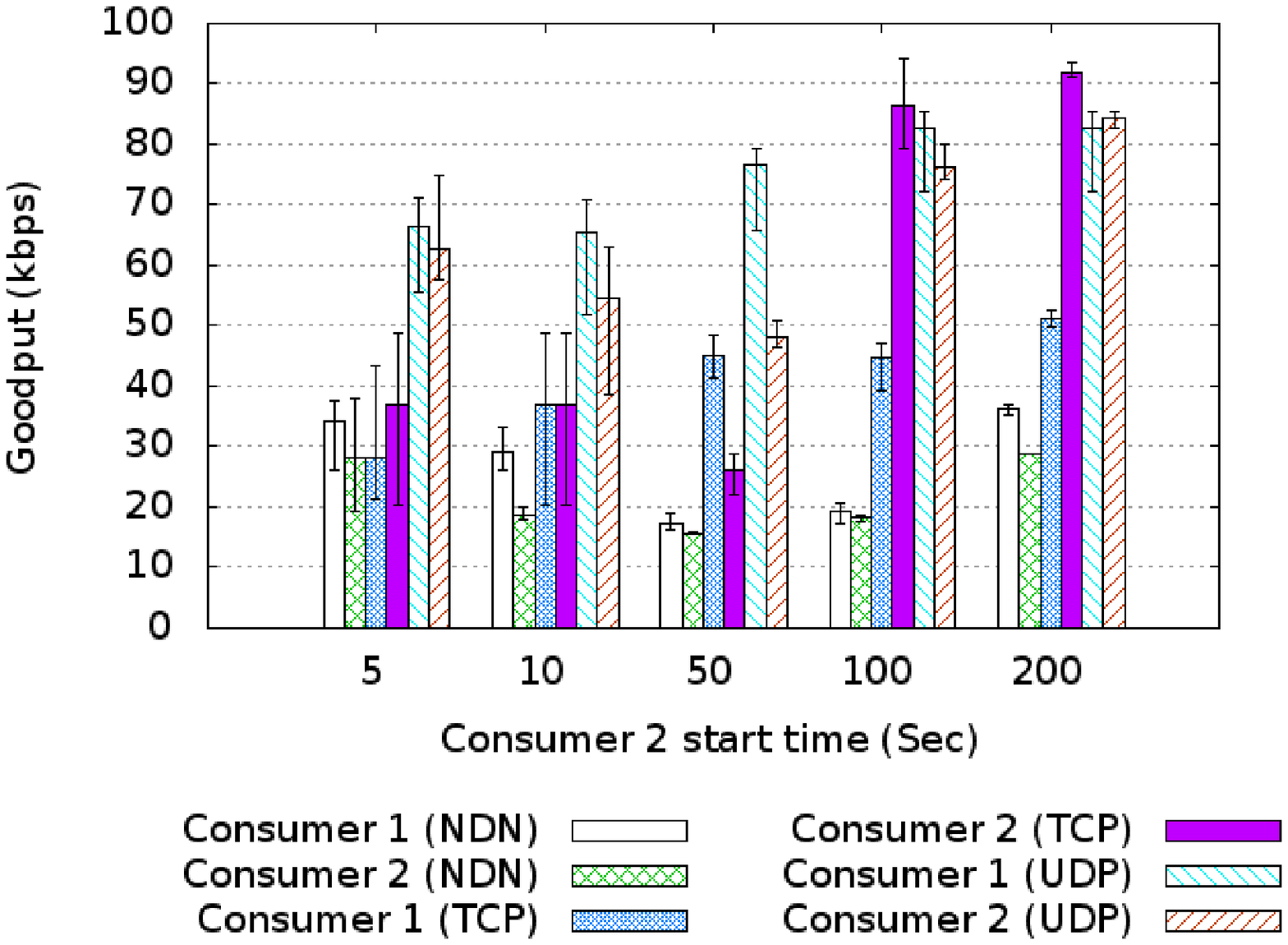} &
    \includegraphics[width=.3\textwidth]{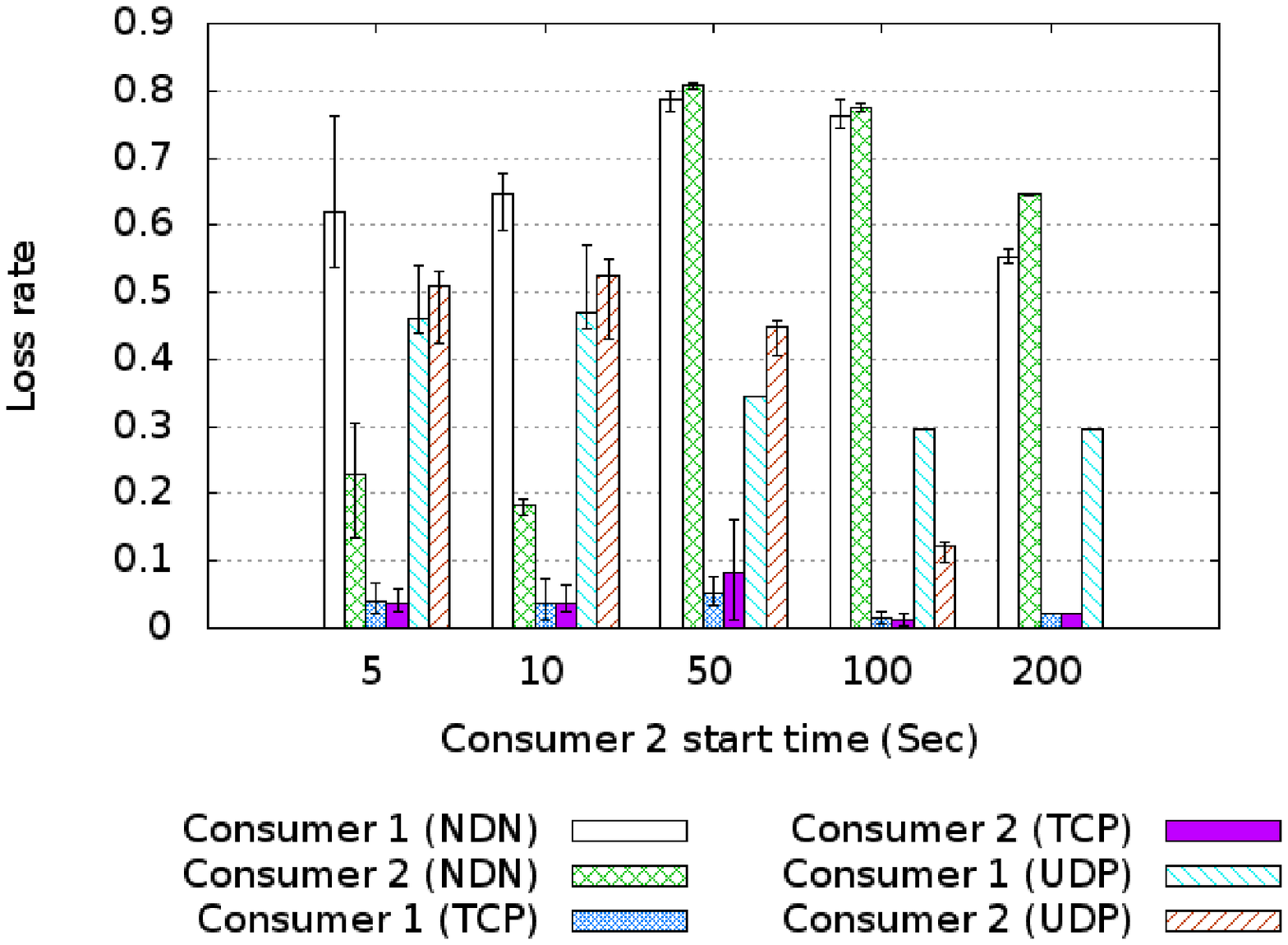}
  \end{tabular}
  \caption{NDN and OLSR goodput and packet loss comparison.}
  \label{fig:Goodput and loss}
\end{figure*}

To compare the performance of NDN with OLSR, we look at the goodput, in Figure \ref{fig:Goodput and loss} (a), along with the packet loss rate, Figure \ref{fig:Goodput and loss} (b).

Intuitively, one would expect that a higher delay would lead to a lower goodput. What we see in the 5 and 10 second cases on NDN is the opposite. The rationale for that is because the goodput is sensitive to the network fragmentation, as we account for the entire duration of the simulation, not just the period where the network is fully connected. 

The difference in the goodput levels in NDN can be explained by network traffic. The network is less congested when Consumer 2 starts at t=200 seconds, and more congested when it starts at t=50 seconds.

It is noticeable that TCP and UDP achieve a higher goodput than NDN in this case. This behavior can be explained by looking at the operation of both routing protocols: while OLSR chooses specific nodes to broadcast (MPR nodes), keeping the wireless channel less congested, the controlled flooding in NDN schedules transmissions at random, but all nodes broadcast at some point, keeping the channel much busier. This reflects on the loss rate plot, where we see that NDN experience more losses than TCP.

From \ref{fig:controlled mobility}, we can observe that the shortest path from the content producer to Consumer 1 is larger than the shortest path to Consumer 2, eight and six hops, respectively. This explains the larger goodput achieved for Consumer 2. We can also see the pattern of network utilization in the goodput cases, where it is less congested in the 200 second case and more congested in the 50 second case.


UDP yields the highest goodput on average, though it is worth noting that both NDN and TCP provide reliability mechanisms, such as retransmissions. This is not the case in UDP, since it transmits at constant rate even if the network is congested. Therefore, the higher goodput in UDP case can be misleading in this case. 

In summary, selecting specific nodes to re-broadcast packets seems to offer a better network utilization, rather than re-broadcasting at random and suppressing overheard packets. This leave us with an open research question that we will explore in the future.

\subsection{Random mobility}
In the scenario where nodes move at random and consumers do not always have a path to the content producer, intermediate nodes should take a store and forward approach, which is naturally done in NDN and in TCP/IP this is implemented via delay-tolerant network (DTN). To study the behavior of NDN in such scenario compared to OLSR and to understand how caches benefit in this case, we start Consumer 2 at different times (e.g. t=0, 10, and 100 seconds).

\begin{figure*}[!t]
\centering
\includegraphics[width=.3\textwidth]{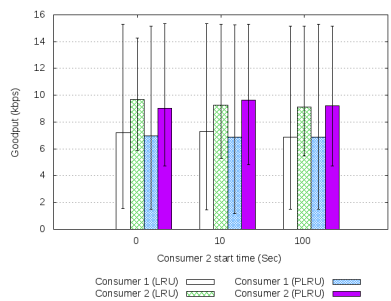}
\caption{NDN and OLSR goodput for random mobility.}
\label{fig:Random goodput}
\end{figure*}

We begin by exploring the behavior of the NDN protocol, depicted in Figure \ref{fig:Random goodput}. As expected, the overall goodput dropped compared to \ref{fig:Goodput and loss} (a), as in this scenario disconnectivity happens more often. Also, we can notice that the variance in all cases is large. This is due to the fact that in some cases in the simulations, consumers have a path to the content source, allowing them to download the file faster, thus the higher goodput. In other cases, this is not true.

All nodes caching with probability $p=0.5$ yield similar goodput results in this case. Consumer 2 achieves 32\% and 35\% higher goodput than Consumer 1 for LRU and PLRU, respectively, due to the fact that its location and movement has more periods of connectivity to the content source, albeit the five different simulation seeds. Because of the periods of disconnectivity, the time at which Consumer 2 starts did not surge effect.

Because our OLSR nodes do not have a store-and-forward capability to support the long periods of disconnectivity, they are not suitable for this scenario. We are studying DTN approaches in this case, though NS-3 does not support delay tolerant networks at this point.

\section{Conclusion} 
\label{Conclusion}
This paper investigates the operability and performance of information-centric networking in MANETs, in a controlled mobility scenario, as well as random walk mobility. We compare NDN -- an instantiation of ICN -- with OLSR -- a representative TCP/IP MANET routing protocol -- when nodes have limited bandwidth and experience intermittent connectivity due to constraints on the mobile devices. For the controlled mobility scenario, OLSR performs better than NDN in terms of data retrieval delay. This can be attributed to the fact that in OLSR, only the MPR nodes will relay the information, whereas in the controlled mobility scheme nodes schedule the rebroadcast using a random timer, leading to a higher loss rate than OLSR. In order to increase the file retrieval delay performance of NDN, we investigate an alternative caching method by deploying probabilistic LRU. In our scenario, PLRU did not show significant improvements on average compared to the plain LRU case. Therefore, we need to further investigate caching schemes to increase caching diversity in the consumer end. In the random mobility scenario, NDN showed that it can still retrieve the file even when a clear path to the content producer is not always available, as opposed to OLSR that needs an end to end path to the source. For future work, we call for a development of a cross-layer deployment of a NDN routing protocol that resembles some aspects of OLSR while maintaining the NDN architecture.


\section{Acknowledgments}
The authors would like to thank Marica Amadeo from the Mediterranean University of Reggio Calabria for providing technical support for this work. The first author is supported by National Council for the Improvement of Higher Education (CAPES), Brazil. Partially funded by ONR contract number N00014-15-C-0122. Views presented are those of authors alone.

%
\bibliographystyle{abbrv}
\bibliography{sigproc}  
%
%

\end{document}